\documentclass[prl,twocolumn,showpacs,floatfix]{revtex4}
\usepackage{amsmath}
\usepackage{verbatim}
\usepackage{graphicx, epsfig}
\usepackage{amssymb}
\usepackage{bm}

\def\be{\begin{equation}}
\def\ee{\end{equation}}
\def\bea{\begin{eqnarray}}
\def\eea{\end{eqnarray}}
\def\bse{\begin{subequations}}
\def\ese{\end{subequations}}

\def\be{\begin{eqnarray}}
\def\ee{\end{eqnarray}}

\newcommand{\ket}[1]{|#1\rangle}
\newcommand{\bra}[1]{\langle #1 |}

\begin{document}

\title{Loss of superfluidity by fermions in the boson Hubbard model on an optical lattice
}
\author{Roman M. Lutchyn$^{1}$, Sumanta Tewari$^{2}$, and
S. {Das Sarma}$^{1}$}
\author{}
\author{}
\affiliation{$^{1}$Condensed Matter Theory Center and Joint Quantum Institute, Department of Physics, University of
Maryland, College Park, MD 20742}
\affiliation{$^{2}$Department of Physics and Astronomy, Clemson University, Clemson, SC 29634}
\date{\today}

\begin{abstract}

 The experimentally observed loss of superfluidity by introducing fermions to the boson Hubbard system on an optical lattice is explained.
We show that the virtual transitions of the bosons to the higher Bloch bands,
coupled with the contact boson-fermion interactions of either sign, result in an effective increase of the
boson on-site repulsion.
If this renormalization of the on-site potential is dominant over the fermion screening of the boson interactions, the Mott insulating lobes of the Bose-Hubbard phase diagram will be enhanced \emph{for either sign of the boson-fermion interactions}. We discuss implications for cold atom experiments where the expansion of the Mott lobes by fermions has been conclusively established.

\end{abstract}

\maketitle

\paragraph{Introduction:}
The intrinsic effect of introducing a gas of fermions on the famous superfluid-insulator (SI) phase diagram \cite{Fisher_PRB89} of the boson Hubbard model is still theoretically unresolved. Experimentally, the situation is quite unambiguous\cite{gunter_PRL06,ospelkaus_PRL06,Bloch08}: the addition of a degenerate gas of spin-polarized fermions to the bosonic superfluid condensate reduces the superfluid coherence, \emph{irrespective of the sign} of the interaction between the bosons and the fermions. This indicates that the condensate, in the presence of interactions with the fermions, gives way to the Mott insulating phase at values of the effective $t/U$ larger than that without the fermions, where $t$ is the nearest neighbor hopping amplitude and $U$ is the on-site interaction of the constituent bosons. The areas of the Mott insulating lobes in the boson Hubbard phase diagram are thus enhanced by the fermions. The earlier experiments \cite{gunter_PRL06,ospelkaus_PRL06} observed the loss of superfluid coherence for fixed attractive boson-fermion interactions, $U_{BF}$, which were larger in magnitude than the boson on-site repulsion itself. Recently this finding has also been confirmed for both attractive and repulsive interspecies interactions in a range of values for $|U_{BF}|$ both smaller and larger than $U$ \cite{Bloch08}. While the problem of the Bose-Fermi mixture is theoretically interesting for the ensuing complexity in the phase diagram \cite{Lewenstein_PRL04, Buchler_PRB04,
pollet_PRA08, Refael_PRB08, Mathey_PRL04, Sengupta_PRA07, Mering_07, Scalettar, Tit, Lutchyn_PRB08}, the main experimental observation -- a suppression of the superfluid coherence by adding fermions -- is contrary to many recent theoretical studies \cite{Buchler_PRB04,pollet_PRA08,Lutchyn_PRB08}; the only notable exception is the numerical work in Ref.~\cite{luhmann-2007}.

  In the theoretical studies of the SI phase diagram of the Bose-Fermi mixture \cite{Buchler_PRB04, Mathey_PRL04, pollet_PRA08,Refael_PRB08,Lutchyn_PRB08}, the central argument involves the screening of the boson on-site repulsive potential by the fermions. The mobile fermions mediate a spatially and temporally non-local \emph{attractive} interaction among the bosons,
   which can be shown to screen and reduce the repulsive interaction $U$ of the boson Hubbard model. Since the on-site
   repulsion tends to localize the bosons on the individual lattice sites (favoring the Mott insulating phases for any
   integer filling factor), any reduction of this interaction, such as that provided by the fermionic screening,
   should expand the area occupied by the superfluid phase in the phase diagram \cite{Buchler_PRB04,pollet_PRA08,Lutchyn_PRB08}.
   As pointed out before, this simple intuitive picture, made rigorous recently both numerically \cite{pollet_PRA08} and
   analytically \cite{Lutchyn_PRB08} on the single band boson Hubbard model, cannot explain the available experiments.
   Here we show that there is an additional effect of the fermion contact interactions of either sign, mediated by virtual
   transitions of the bosons to the higher Bloch bands, which leads to an effective \emph{increase} of the
   boson on-site interactions. There is some numerical evidence of this effect \cite{luhmann-2007} for the case
   of attractive interspecies interactions only.
  In this Letter, we treat the above two disparate effects within a unified analytical framework. We believe that this theory, which includes the idea of virtual transitions to the higher boson Bloch bands,
  provides an explanation for the loss of bosonic superfluid coherence by introducing fermions (irrespective of the sign of the interspecies interactions),
  seen in the recent cold atom experiments. Our conclusion is that multi-band effects, neglected in the previous studies~\cite{Lewenstein_PRL04, Buchler_PRB04,
pollet_PRA08, Refael_PRB08, Mathey_PRL04, Sengupta_PRA07, Mering_07, Scalettar, Tit, Lutchyn_PRB08}, are important to describe the physics of Bose-Fermi mixtures. 

  To get a unified description of the effects of the fermions, we start with the multi-band boson Hubbard model where the two lowest-lying boson Bloch bands are separated by an energy gap $\Omega \gg t, \mu, U$, where $\mu$ is the boson chemical potential. We assume that the fermions are coupled to the bosons in the two lowest-lying Bloch bands with contact interactions. The trivial effects of the presence of the higher boson bands are renormalizations of the bare bosonic parameters $t, \mu, U$, which are present even without the fermions (and hence, will be mostly suppressed here). However, most importantly, we find that the virtual transitions of the bosons to the higher Bloch bands give rise to a new type of on-site \emph{repulsive} interaction among the bosons mediated by the fermions. This interaction, which is nonlinear in the boson-fermion scattering length $a_{BF}$, and remains repulsive for either sign of $a_{BF}$,
 tends to hinder the flow of the bosons on the optical lattice. Thus, this interaction moves the Mott insulating transitions
  to shallower lattice depths for any boson integer filling factor. Including this new interaction and the usual fermion-mediated screening
  interaction \cite{Buchler_PRB04,pollet_PRA08, Refael_PRB08, Lutchyn_PRB08} in a unified framework \cite{Lutchyn_PRB08},
  we find that the question of the overall shift of the phase diagram
 is a quantitative one arising from a competition between attractive and repulsive terms: if the fermion-induced higher-band renormalization of the on-site potential is  dominant, as may be the case in the recent experiments, superfluid coherence will be reduced by introducing the fermions.
\paragraph{Model:}
We begin with the following second quantized Hamiltonian
describing the bosons and spin-polarized fermions interacting with each other through a contact interaction in an optical lattice:
\vspace*{-.05 in}
\begin{align}\label{eq:GP}
H&\!=\!\int\!d^3 \bm r \Phi^\dag (\bm r)\!\left[ H^B_0\!+\!\frac{g_{BB}}{2}\Phi^\dag (\bm r)\Phi (\bm r)\right] \Phi (\bm r)+\\
&\int \! d^3 \bm r \Psi^\dag (\bm r) H^F_0 \Psi (\bm r)\!+\!\frac{g_{BF}}{2}\int \!d^d \bm r \Phi^\dag (\bm r)\Phi (\bm r) \Psi^\dag (\bm r)  \Psi (\bm r)\nonumber.
\end{align}
Here, $\Phi( \bm r)$ and $\Psi( \bm r)$ are the field operators for the bosonic and the fermionic atoms, respectively, the single particle Hamiltonians $H^{B/F}_0={\hat T}_{B/F}+V_{\rm lat}(\bm r)$, with $ \hat T_{B/F}$ representing the kinetic energy for the bosons/fermions and $V_{\rm lat}(\bm r)$  denoting the lattice potential, $V_{\rm lat}(\bm r)=\sum_{j=1}^3V_0\sin^2(\pi \frac{r_j}{a})$ with $a$ as the lattice spacing. The interaction coupling constants are given by $g_{BB}=\frac{ 4\pi a_{BB}}{m_B}$, $g_{BF}=\frac{ 4\pi a_{BF}}{m_{\rm red} }$, where $m_B$ is the mass of a bosonic atom, $m_{\rm red}$ is the boson-fermion reduced mass and $a_{BB/BF}$ are the boson-boson and boson-fermion scattering lengths, respectively.

We now expand the field operators in the Wannier function basis, $\Phi( \bm r)=\sum_{i,\alpha} b_{i,\alpha} w_{\alpha}(\bm r\!-\!\bm r_i)$, and $\Psi( \bm r)=\sum_{i,\alpha} u_{\alpha}(\bm r - \bm r_i) c_{i,\alpha}$, where the operators $b_{i,\alpha}$ and $c_{i,\alpha}$ annihilate the bosons and the fermions at a site $i$ in a band $\alpha$, respectively. Substituting the above expressions for the field operators into Eq.~(\ref{eq:GP}), one obtains the full multi-band model \cite{Scarola_PRL05, Isacsson_PRA05, Liu_PRL06} for the Bose-Fermi system. Instead of treating the complexity of the full multi-band Hamiltonian, we consider, for simplicity, a two-band model for the bosons with the fermions being in a single band, and keep the largest band-mixing terms in the bosonic part of the Hamiltonian. This effective two-band model for the bosons, which can be justified for large interband energy separation $\Omega=\sqrt{4E_RV_0}$, where $E_R=\pi^2/2m_B a^2$ is the recoil energy, captures the essential physics involving the virtual transitions of the bosons to the higher Bloch bands.
 The fermion Wannier wavefunctions $u_{\alpha}(\bm r-\bm r_i)$ are chosen using the mean-field one-body Hamiltonian for the fermions, $\hat T_F+V_F(\bm r)$, where the effective potential $V_F(\bm r)=V_{\rm lat}(\bm r)+\frac{g_{BF}}{2}\rho_B(\bm r)$. Here, $\rho_B(\bm r)=n_0 |w_{i,1}(\bm r)|^2$ with $\rho_B(\bm r)$ and $n_0$ being the average boson density per site and average boson number per site, respectfully. Thus, the shapes of these functions within a unit cell, which will be important later to determine the sign of the fermion renormalization of the on-site bosonic potential (see the discussion after Eq.~(\ref{eq:p})), depend on the sign of the interspecies interactions (sign of $a_{BF}$). In the rest of the paper we will study the two-band Bose-Fermi model defined by the Hamiltonian,
\begin{align}\label{eq:Bose-Hubbard-two-band}
\!H&=H_l+H'_l+H_t+H_{BF}+H_F\\
\!H_l\!&\!=\!\sum_{i,\alpha}\!\! \left[\varepsilon_{\alpha}\hat{n}_{i,\alpha}\!+\!\frac{U^{\alpha,\alpha}}{2}\hat{n}_{i,\alpha}(\hat{n}_{i,\alpha}\!-\!1)\!\right]\!+\!\!\sum_{i,\alpha > \alpha'}\!\!\!2U^{\alpha,\alpha'}\!\hat{n}_{i,\alpha}\hat{n}_{i,\alpha'}\! \nonumber\\
\!H'_l \!&\!=\!\sum_{i,\alpha > \alpha'}\!\!\!\frac{U^{\alpha,\alpha'}}{2}\left[b^{\dag}_{i,\alpha'}b^{\dag}_{i,\alpha'}b_{i,\alpha}b_{i,\alpha}\!+\!b^{\dag}_{i,\alpha}b^{\dag}_{i,\alpha}b_{i,\alpha'}b_{i,\alpha'}\right]\nonumber\\
\!H_t\!&\!=\!-\!\!\!\sum_{<ij>,\alpha}\!\!t^{\alpha}\!\!\left[b^{\dag}_{i,\alpha} b_{j,\alpha}
\!+\! \rm H.c.\!\right]\!\!; H_{BF}\!\!=\!\sum_{i,\alpha} U^{\alpha}_{FB}[\hat n_{i,\alpha}\!-\!\langle \hat n_{i,\alpha}\!\rangle\!] \hat n^F_{i}\nonumber\\
\!H_F\!&=\!\sum_{<ij>}\left[\epsilon_0\hat{n}^F_{i}\delta_{ij}-t_F\left(c^{\dag}_{i} c_{j}
\!+\!H.c.\right)\right].\nonumber
\end{align}
 Here, $\alpha=1,2$, the energies $\varepsilon_{\alpha}=\bra{w_{i,\alpha}}H^B_0\ket{w_{i,\alpha}}-\mu$, the matrix elements $U^{\alpha,\alpha'}=g_{BB}\bra{w_{i,\alpha};w_{i,\alpha'}}\! \ket{w_{i,\alpha};w_{i,\alpha'}}$ and $U^{\alpha}_{FB}=g_{BF}/2\bra{w_{i,\alpha};u_{i}}\!\! \ket{w_{i,\alpha};u_{i}}$. The fermion energy and hopping are given by $\epsilon_0=\bra{u_{i}} \hat{T}_F+V_F(\bm r)\ket{u_{i}}-\mu$  and $t_F=-\bra{u_{i}} \hat{T}_F+V_F(\bm r)\ket{u_{j}}$. Note that the piece $H'_l$ in Eq.~(\ref{eq:Bose-Hubbard-two-band}), which corresponds to scattering of two bosons between the first and the second Bloch bands, leads to band mixing~\cite{higherband2}.  As we show below, these band-mixing terms, coupled with the fermion contact interactions, renormalize the local repulsive interaction between the constituent bosons in the low energy subspace.

 \paragraph{Schrieffer-Wolff transformation:} In order to reveal the nature of the renormalization, one needs to diagonalize the full Hamiltonian and project it on the subspace of the lower band only.
 Assuming that $\Omega$ is the largest energy scale in the problem, we perform the Schrieffer-Wolff canonical transformation~\cite{Schrieffer-Wolff} on the Hamiltonian $H$ to decouple the diagonal (boson number-conserving) and the non-diagonal (band-mixing) pieces of the Hamiltonian to a given order in $1/\Omega$:
\begin{equation}\label{eq:canonical}
H_{\rm eff}= e^{S}He^{-S}=H+[S,H]+\frac 12 [S,[S,H]]+...
\end{equation}
 To look for the right unitary transformation operator $S$, we separate the Hamiltonian in Eq.~(\ref{eq:Bose-Hubbard-two-band}) into three parts: $H_0=\Omega \sum_i \hat n_{i,2}$, $H_1=H-H_0-H_2$ and $H_2=H'_l$. To the zeroth order in $1/\Omega$ the operator $S^{(1)}$ is given by
\begin{align}
\!\!S^{(1)}\!=\!\!\!\sum_{i,\alpha>\alpha'} \! \frac{U^{\alpha,\alpha'}}{4\Omega}\! \left(\!b^{\dag}_{i,\alpha}b^{\dag}_{i,\alpha}b_{i,\alpha'}b_{i,\alpha'}\!-\!b^{\dag}_{i,\alpha'}b^{\dag}_{i,\alpha'}b_{i,\alpha}b_{i,\alpha}\!\right)\!\!.
\end{align}
One can check that $S^{(1)}$ satisfies $[S^{(1)}, H_0]=-H_2$, so, to the zeroth order in $1/\Omega$, the effective Hamiltonian is simply the sum of $H_0$ and $H_1$. The next order operators $S^{(2)}$ and $S^{(3)}$ can be found recursively, \emph{e.g.}, $S^{(2)}=[H_2, H_1]/4\Omega^2$.  Using the canonical transformation defined in Eq.~(\ref{eq:canonical}) with $S=S^{(1)}+S^{(2)}+S^{(3)}$, we decouple $H$ to the order $1/\Omega^2$: $H_{\rm eff}=H_0+H_1+\frac 12 [S^{(1)},H_2]+\frac 12 [S^{(2)},H_2]$. Then, by projecting $H_{\rm eff}$ on the lowest Bloch band, and omitting the band index, we arrive at the following low-energy Hamiltonian to this order:
\begin{align}\label{eq:Heff}
 \tilde H_{\rm eff}\!&\!=\!\sum_i \!\left[\!\frac{1}{2}\!\left(\!\tilde U\!-\! \frac{p U_{FB}}{2}\hat n^F_i\! \left(\!\frac{U^{1,2}}{\Omega}\!\right)^2\! \right)\! \hat{n}_{i}(\hat{n}_{i}\!-\!1)\!-\!\tilde \mu \hat{n}_{i}\!\right]\!+\!\tilde H_t \nonumber \\
\!&+\!\sum_{i}U_{FB}(\hat n_{i}-\langle \hat n_{i}\rangle ) \hat n^F_i + H_F.
\end{align}
The tilde indicates the renormalization of the original boson Hubbard parameters $U$, $t$ and $\mu$ due to processes involving virtual transitions of the bosons to the higher Bloch band, \emph{e.g.}, $\tilde U=U-(U^{1,2})^2/\Omega+...$. These renormalizations are also present in a pure bosonic system and are independent of $U_{FB}$. Most importantly, however, the contact interactions with the fermions provide an additional renormalization to the boson on-site potential $\tilde{U}$ which is linear in $U_{FB}$.

The correction to the boson-boson repulsion
due to the fermions corresponds to a virtual process when two bosons are excited from the first to the second Bloch band.
The probability for such processes to occur (\emph{i.e.},
 the fraction of time the system dwells in such a virtual state) is proportional to $(U^{1,2}/\Omega)^2 n_i (n_i-1)$.
 While in the virtual state, the interaction energy between the bosons and the fermions
 (defined by $H_{BF}$ in Eq.~(\ref{eq:Bose-Hubbard-two-band})) changes by $2 (U^{(2)}_{BF}-U^{(1)}_{BF}) n_{Fi}$.
 Thus, by combining the above two terms and introducing the dimensionless parameter $p$ (to account for the change in the interaction energy),
 \begin{align}
 p=1\!-\!\bra{w_{i,2};u_{i}}\!\! \ket{w_{i,2};u_{i}}/\bra{w_{i,1};u_{i}}\!\! \ket{w_{i,1};u_{i}},
 \label{eq:p}
 \end{align}
 one recovers the correction to the boson-boson repulsion given in Eq.~(\ref{eq:Heff}).
 For attractive boson-fermion interactions, the fermionic wavefunction $u(\bm r)$ is peaked at the center. Thus, the overlap of $u_i$ with the boson Wannier function in the second Bloch band, $w_{i,2}$, is smaller than its overlap with $w_{i,1}$. Therefore, for negative $U_{BF}$, $p$ is positive and ${\cal{O}}(1)$. However, $p$ changes sign for repulsive $U_{BF}$, since, in this case, the two species of atoms maximize the distance between them
 (\emph{i.e.}, the fermion density is suppressed at the center of the unit cell), resulting in the numerator in the second term in Eq.~(\ref{eq:p}) exceeding the denominator.
 From Eq.~(\ref{eq:Heff}), notice that the sign of the renormalization to $\tilde{U}$ is determined by sgn$(pU_{FB})$.
 Therefore, it remains repulsive and tends to suppress the superfluid phase \emph{for either sign of $U_{BF}$}.

 \paragraph{Shift of the phase diagram:}
To the lowest order in $U_{FB}$, the on-site interaction energy and the chemical potential of the bosons are modified as $\tilde{U}\rightarrow U'=\tilde{U}-pU_{FB}\left(U^{1,2}/\sqrt{2}\Omega\right)^2 n^0_{Fi}$ and $\tilde{\mu}\rightarrow\mu'=\tilde{\mu} -U_{FB}n^0_{Fi}$, with $n^0_{Fi}$  the average density of the fermions.
Since the fermions appear only at the quadratic order in Eq.~(\ref{eq:Heff}), integrating them out leads to the effective imaginary-time action~\cite{derivation}
\begin{align}\label{eq:eff_action}
\!\!&\!\!S_{\rm eff}(b^{*}, b)=\sum_i\!\int_{0}^{\beta}\!\! d\tau \left[b^{*}_i \partial_{\tau} b_i+ \frac{U'}{2}\hat{n}_{i}(\hat{n}_{i}\!-\!1)
\!-\! \mu' \hat{n}_{i}\!\right]\\
\!&\!-\! \int_{0}^{\beta}\!\! d\tau\! \tilde H_t \!-\!\sum_{ij}\!\! \int_0^{\beta}\! d \tau_1\! \int_0^{\beta} d \tau_2
n_{i}(\tau_1)M_{ij}(\tau_1\!-\!\tau_2) n_{j}(\tau_2)\nonumber
\end{align}
  From the last term in Eq.~(\ref{eq:eff_action}), it is clear that the fermions induce another boson-boson interaction with the spatially and temporally non-local kernel (to the order $U^2_{FB}/\Delta$) $M_{ij}(\tau_1-\tau_2)=U^2_{FB}\langle \delta n_{Fi}(\tau_1) \delta n_{Fj}(\tau_2)\rangle/2$. Here, $\delta n_{Fi}=n_{Fi}-n^0_{Fi}$ and the parameter $\Delta$ is proportional to the inverse of the density of states at the Fermi level $\nu_F$,  $\Delta = 1/\nu_F a^3$. This contribution describes
the screening of the bosonic repulsive interactions by the fermions, which effectively reduces $U$, leading
to the suppression of the Mott insulating phase~\cite{Buchler_PRB04,pollet_PRA08,Lutchyn_PRB08}.

\begin{figure}
\centering
\includegraphics[width=0.99\linewidth]{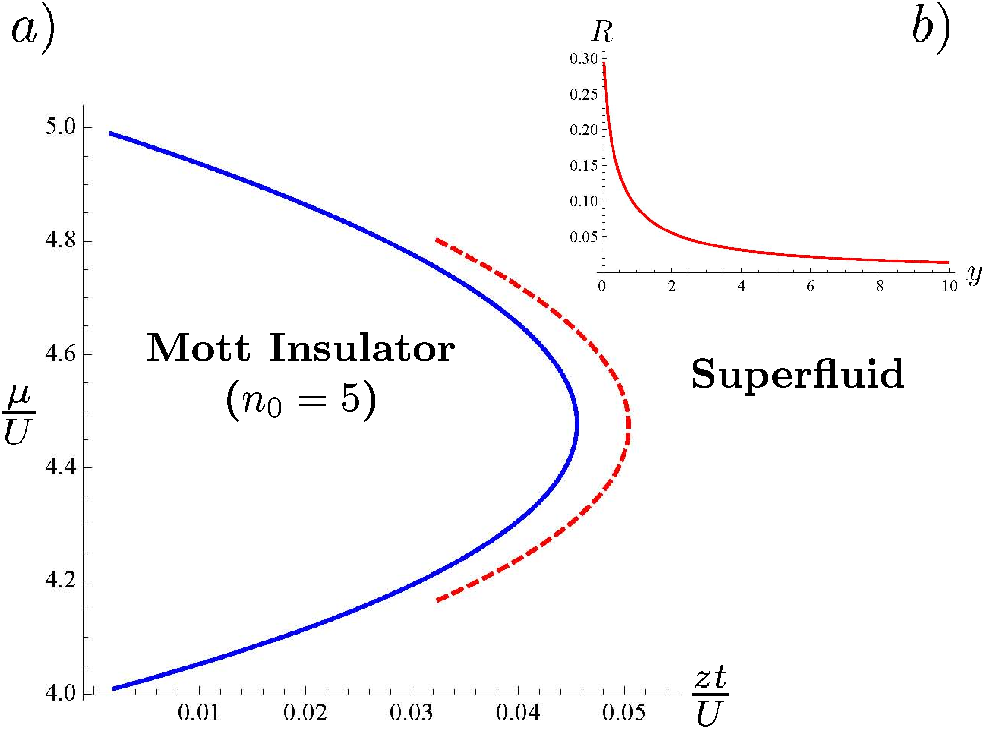}
  \caption{(Color online) (a) Main panel: Phase Boundary of the boson Hubbard model with and without the fermions for the boson density $n_0=5$. Solid line describes the
 insulator-superfluid phase boundary without the fermions. The dashed line corresponds to the phase boundary with the fermions present in the region of the validity of the perturbation series~\cite{breakdown}.
Here we used $U_{FB}\approx - U_{BB}$, $U^{1,2}/\Omega \approx 0.3$, $\Delta/U_{BB}=40$ and $n^0_{Fi}=0.75$. (b) Inset: The dependence of the function $R(y)$ on its argument. } \label{fig:phase_boundary}
\end{figure}

The effect of the above two competing contributions on the phase diagram can be calculated analytically. We first need to calculate the boson on-site Green's function \cite{Sachdev_book} for the action in Eq.~(\ref{eq:eff_action})
at zero frequency \cite{Lutchyn_PRB08},
\begin{align}\label{eq:Green's}
 \!&\!\!{{G}}_i(0)\!=\\
 \!&\!-\!\frac{n_0\!+\!1}{\delta E_p}\!\left[\!1\!+\!\frac{p\, U_{FB} n_0 n^0_{Fi}}{2\delta E_p}\!\!\left(\!\frac{U^{1,2}}{\Omega}\!\right)^2\!+\!\frac{U_{FB}^2}{\Delta \delta E_p}\!R\!\left(\!\frac{\delta E_p}{4E_F} \! \right) \! \right]\nonumber\\
 \!&\!-\!\frac{n_0}{\delta E_h}\!\left[\!1\!+\!\frac{p\, U_{FB} (1-n_0) n^0_{Fi}}{2\delta E_h}\!\!\left(\!\frac{U^{1,2}}{\Omega}\!\right)^2\!+\!\frac{U_{FB}^2}{\Delta \delta E_h}\!R\!\left(\!\frac{\delta E_h}{4E_F} \! \right) \! \right]\!.\nonumber
\end{align}
Here $\delta E_p$ and $\delta E_h$ are the particle and the hole excitation energies: $ \delta E_p={U' n_0 -\mu'}$ and $\delta
E_h={\mu' -U' (n_0 -1)}$, $n_0$ is the number of bosons per site minimizing the ground state energy.
The dimensionless function $R(y)=-\frac{3}{\pi} \int_0^1x^2dx\int_0^\infty d\nu \frac{y}{y^2+\nu^2}\Pi(x,\nu)/\nu_F$,
where $\Pi(\bm q,\Omega_n)$ is the three-dimensional fermion polarization function \cite{Fetter_book}. The dependence of $R(y)$ on its argument is plotted in Fig.~\ref{fig:phase_boundary}.  The mean field
SI phase boundary can be obtained by solving the equation
$1/z \tilde{t}+\int_{-\beta}^{\beta} d\tau {{G}}_i(\tau)=0$,
where $z$ is the lattice coordination number~\cite{Sachdev_book}.

 To the linear order in $U_{FB}$, the correction to the phase boundary is influenced only by the enhancement of the on-site repulsion given in Eq.~(\ref{eq:Heff}):
\begin{align}
\frac{\delta t^{(1)}}{\tilde U}&\!\equiv\! \frac{t(a_{BF})\!-\!t(a_{BF}\!=\!0)}{\tilde U}\!=\! - \frac{  U_{FB} p}{2 \tilde U} \!\left(\frac{{U^{1, 2}}}{\Omega}\right)^2 \! n^0_F\\  &\times\left[\frac{[1+2(\tilde \mu/ \tilde U)^2-2(\tilde \mu/\tilde U)(n_0-1)-n_0] n_0}{[1+(\tilde \mu/\tilde U)]^2}\right]\nonumber
\end{align}
Notice that for either attractive or repulsive $U_{BF}$, the product of $U_{BF}$ and $p$ remains negative. Therefore, the shift $\delta t^{(1)}$
is towards shallower lattices, indicating an expansion of the Mott-insulating lobes.
The shift of the phase boundary due to the fermion-mediated screening, which manifests itself only in the second order in $U_{FB}$, is given by,
\begin{align}\label{eq:Green's_dyn}
\!\frac{\delta t^{(2)}}{\tilde U}\!&\!=\!- \frac{U_{FB}^2}{\Delta \tilde U \!\left(\!1\!+\!\frac{\tilde \mu}{\tilde U}\right)^2}\! \left \{ \!{n_0\! \left(\!n_0\!-\! \frac{\tilde \mu}{\tilde U}\right)^2}\!\! R\!\left(\!\!\frac{\tilde U}{4E_F}\!\left[\frac {\tilde \mu} {\tilde U} \!-\!n_0\!+\!1\!\right]\! \right) \right. \nonumber \\
& \left. \!+{(n_0\!+\!1)\!\left(\!\frac{\tilde \mu} {\tilde U}\!-\!(n_0\!-\!1)\!\right)^2}\! R\!\left(\!\frac{\tilde U}{4E_F}\!\left[\!n_0\!-\!\frac {\tilde \mu} {\tilde U} \right]\right)  \! \right \}\!. 
\end{align}
As expected, the fermion-mediated screening enhances the area occupied by the superfluid phase.
Notice that the small perturbation parameters used in the above calculations are $\frac{U^{12}}{\Omega}$ and $\frac{U_{FB}}{\Delta}$, respectively. Therefore, they can be parametrically valid for both $|U_{BF}|$ smaller as well as larger than the bare $U$.
At the tip of the Mott lobes, the ratio of the two contributions for $n_0 \gg 1$ is given by,
\begin{align}
\left |\frac{\delta t^{(1)}}{\delta t^{(2)}}\right|\sim\frac{|p| n_F^0 \Delta}{|U_{FB}| n_0 R(\frac{\tilde U}{8E_F})}\left(\frac{U^{12}}{\Omega}\right)^2 
\end{align}
 Since the two competing contributions depend on different {\it independent} parameters, the sign of the phase boundary shift is a quantitative question.
  For example, for $(\frac{U^{12}}{\Omega})^2 \sim \frac{|U_{FB}|}{\Delta}$, the two effects are comparable. In general, for $\sqrt{\frac{|U_{FB}|}{\Delta}} <\frac{U^{12}}{\Omega} $, the superfluid state is suppressed for either sign of the interspecies interaction, as shown in the figure~\ref{fig:phase_boundary}.
\paragraph{Conclusion:} We show that the virtual transitions of bosons to
 the higher Bloch bands in an optical lattice, coupled with contact interactions with a degenerate gas of fermions,
  generate a new renormalization of the interactions in the boson Hubbard model. For either sign of the coupling between the fermions and
the bosons, this renormalization enhances the boson on-site
 repulsion, and thus favors the Mott insulating phase.
  If this effect is dominant over the usual fermion mediated screening, the superfluid coherence of the Bose-Hubbard system will be suppressed by the fermions, as has been observed in recent experiments~\cite{gunter_PRL06,ospelkaus_PRL06,Bloch08}. Finally, we emphasize that promoting bosons to higher Bloch bands can lead to new types of heteronuclear interactions which are important for the realistic description of heteronuclear mixtures in cold-atom experiments.

We thank L.~Cywinski, T.~Porto and W.~Phillips for stimulating discussions. This work is supported by ARO-DARPA and the Clemson University start-up funds.


\begin{thebibliography}{17}
\bibitem{derivation}
When deriving the effective action $S_{\rm eff}$ we neglected the term proportional to $U^2_{FB}/\Omega^4$ since it is much smaller than the last term in Eq.~(\ref{eq:eff_action}).
\bibitem{breakdown} Near the degeneracy points~[\onlinecite{Sachdev_book}] where $\delta
  E_{p/h}$ is of the order~of $U_{FB}^2/\Delta$ the perturbation theory breaks down.
\end{thebibliography}
\end{document}